\documentclass{mem}
\usepackage{natbib}\usepackage{txfonts}\usepackage{balance}
\usepackage{graphicx}
\usepackage[a4paper]{hyperref}
\idline{75}{282}
\begin{document}
\def\teff{$T\rm_{eff }$}
\def\kms{$\mathrm {km s}^{-1}$}
\def\ms{$\mathrm {m s}^{-1}$}
\def\bz{$\langle B_z\rangle$}
\def\ii{\,{\sc ii}} \def\iii{\,{\sc iii}}

\title{
Spectroscopy of roAp star pulsation: HD\,24712 }

   \subtitle{}

\author{
M. \,Sachkov\inst{1}, T. \,Ryabchikova\inst{1,2}, S.
\,Bagnulo\inst{3}, I. \,Ilyin\inst{4}, T. \,Kallinger\inst{2},\\ 
O. \,Kochukhov\inst{5}, F. \,Leone\inst{6}, G. \,Lo Curto\inst{3},
T. \,L\"uftinger\inst{2}, D. \,Lyashko\inst{7}, A.
\,Magazzu\inst{8},\\  H. \,Saio\inst{9}, \and W.W.
\,Weiss\inst{2} }

 \offprints{M. Sachkov, \email{msachkov@inasan.ru}}

\institute{ Institute of Astronomy, Russian Academy of Science, 48
Pyatnitskaya str., 119017 Moscow, Russia
\and Institute for Astronomy,
University of Vienna, T\"urkenschanzstrasse 17, A-1180 Vienna,
Austria \and European Southern Observatory, Casilla 19001,
Santiago 19, Chile \and Astrothysikalisches Institut Potsdam, An
der Sternwarte 16, D-14482 Potsdam, Germany \and Department of
Astronomy and Space Physics, Uppsala University Box 515, SE-751 20
Uppsala, Sweden \and INAF - Osservatorio Astrofisico di Catania, Via S.
Sofia 78, 95123 Catania, Italy \and Tavrian National University,
Simferopol, Ukraine \and INAF - Telescopio Nazionale Galileo, PO
Box 565, 38700 Santa Cruz de La Palma, Spain \and Astronomical
Institute, Tohoku University, Sendai, Miyagi 980-8578, Japan  \\
}

\authorrunning{Sachkov et al.}

\titlerunning{Pulsations in roAp star HD\,24712}

\abstract{ We present results of the radial velocity (RV) analysis
of spectroscopic time-series observations of the roAp star
HD\,24712 (HR\,1217) which were carried out simultaneously with
the Canadian MOST mini-satellite photometry. Only lines of the
rare-earth elements (REE) show substantial amplitudes of RV
pulsations. Based on new Zeeman measurements we found different
shapes of the magnetic curves derived by using Fe-peak and REE
separately. Frequency analysis of the spectroscopic data showed
that the highest amplitude frequencies are the same in photometry
and spectroscopy. Photometric and spectroscopic pulsation curves
are shifted in phase, and the phase shift depends on the atomic
species. The observed distribution of RV pulsation amplitudes
and phases with the optical depth as well as the observed phase
lag between luminosity and radius variations are explained
satisfactorily by the model of nonadiabatic nonradial pulsations
of a magnetic star.
  }

\maketitle{}

\section{Introduction}

HD\,24712 is one of the spectroscopically best-studied roAp star
(\citealt{bal_z}, \citealt{sach04}, \citealt{sach04b},
\citealt{mkrt05}). Because of the known changes of the pulsational
characteristics with time, a simultaneous spectroscopic (magnetic)
and photometric monitoring during the whole rotation period is
needed to avoid ambiguity when interpreting the relation between
spectroscopic and photometric evidence for pulsation. Continuous
multi-site or space photometry of multiperiodic stars provides 
a reliable frequency
solution needed to derive information
about the stellar interior, while radial velocity (RV) analyses
allow for a detailed study of the pulsation phenomena in a stellar
atmosphere.

\section{Spectroscopic observations}

\begin{figure*}[t]
\includegraphics[width=100mm]{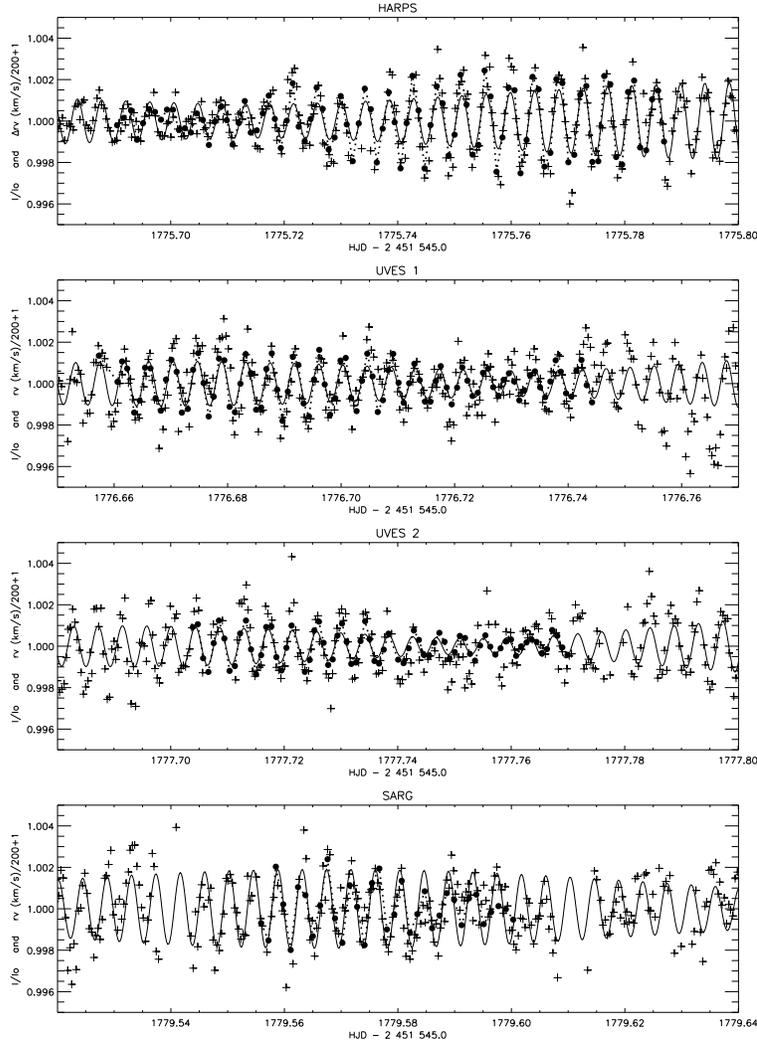}
\parbox[b]{32mm}{
\caption{ \footnotesize Normalized RV variations for Pr\,\iii\
(filled circles connected by dotted line) are compared with the
simultaneous MOST photometry (pluses). Black solid line represents
theoretical curve generated with the frequency solution for the
whole MOST observing run, kindly provided by the MOST team. Phase
shift between photometric and spectroscopic variations is given on
top of the figure. } \label{shift}}

\end{figure*}

Simultaneously with the Canadian minisatellite MOST, which
monitored HD\,24712 from November 6 to December 5, 2004,
we have obtained time-series high-resolution spectroscopic observations.
They were carried out at ESO during November
10/11, 2004 (HARPS -- 92 spectra, 60 sec time resolution, S/N=120),
and, because of the unique coincidence with the
space photometry, Director's Discretionary Time (DDT-274.D-5011) was
granted for 11/12 and 12/13 (UVES -- 92 \& 73 spectra, 50 sec, S/N=300).
Furthermore, 35 spectra were obtained on November 13/14, 2004 with SARG at TNG.
The spectroscopy covers the rotation phases of 0.867, 0.944, 0.028, and 0.176
which coincide with magnetic and pulsation amplitude maximum according 
to the most recent ephemeris \citep{RW05}.
All spectra were
reduced and normalized to the continuum level either with MIDAS
and IRAF or with a routine specially developed by one of us (DL)
for a fast reduction of time-series observations.

Zeeman observations were obtained at NOT with the SOFIN
spectrograph during 13 consecutive nights in November 2003 and
hence cover a complete rotation period.

\section{Radial velocity and magnetic field analysis}

Radial velocities of more that 500 unblended spectral lines were
measured in the spectral region from 3900 to 6800~\AA. For the
first time RV measurements were made before and after the Balmer
Jump (BJ) using the UVES spectra. About 1/3 of all pulsating lines
could not be identified, but according to their pulsation
characteristics they should belong to rare-earth elements (REE).
We confirmed our previous results (\citealt{sach04b}) that only
REE lines and the H$\alpha$ core show large pulsation amplitudes
(150 -- 400 \ms), while spectral lines of the other elements are
constant (Mg, Si, Ca, Fe-peak) or only weakly pulsating (the very
cores of the resonance lines of Ca\ii\ and Sr\ii\ and the H$\beta$
line core). No difference in the pulsation signature is found for
the lines of the same element/ion on both sides of the BJ. We
confirm the same phase shifts between RV variations in the lines
of different elements/ions found earlier by \citet{sach04b},
indicating stability of the pulsation mechanism in the atmosphere
of HD\,24712 at least during  the last years.

Our four nights of spectroscopic monitoring are not sufficient for
a detailed frequency analysis despite the high accuracy of the RV
measurements. We may only conclude that the 3 highest amplitude
frequencies are the same in spectroscopy and MOST photometry which
is also supported by a direct comparison (Fig.\,\ref{shift}). Photometry and spectroscopy are
shifted in phase, depending on the atomic species and on the depth
of the line forming region. For the first time it was possible to
derive a phase lag between luminosity and radius variations at
different levels in the atmosphere. The luminosity maximum occurs
0.58\,$P_{\rm puls}$ (3.7 radian) after the minimum radius of a
layer at $\log \tau_{5000}$\,=\,-3.7.

\begin{figure}[h]
\includegraphics[width=5.5cm,angle=-90]{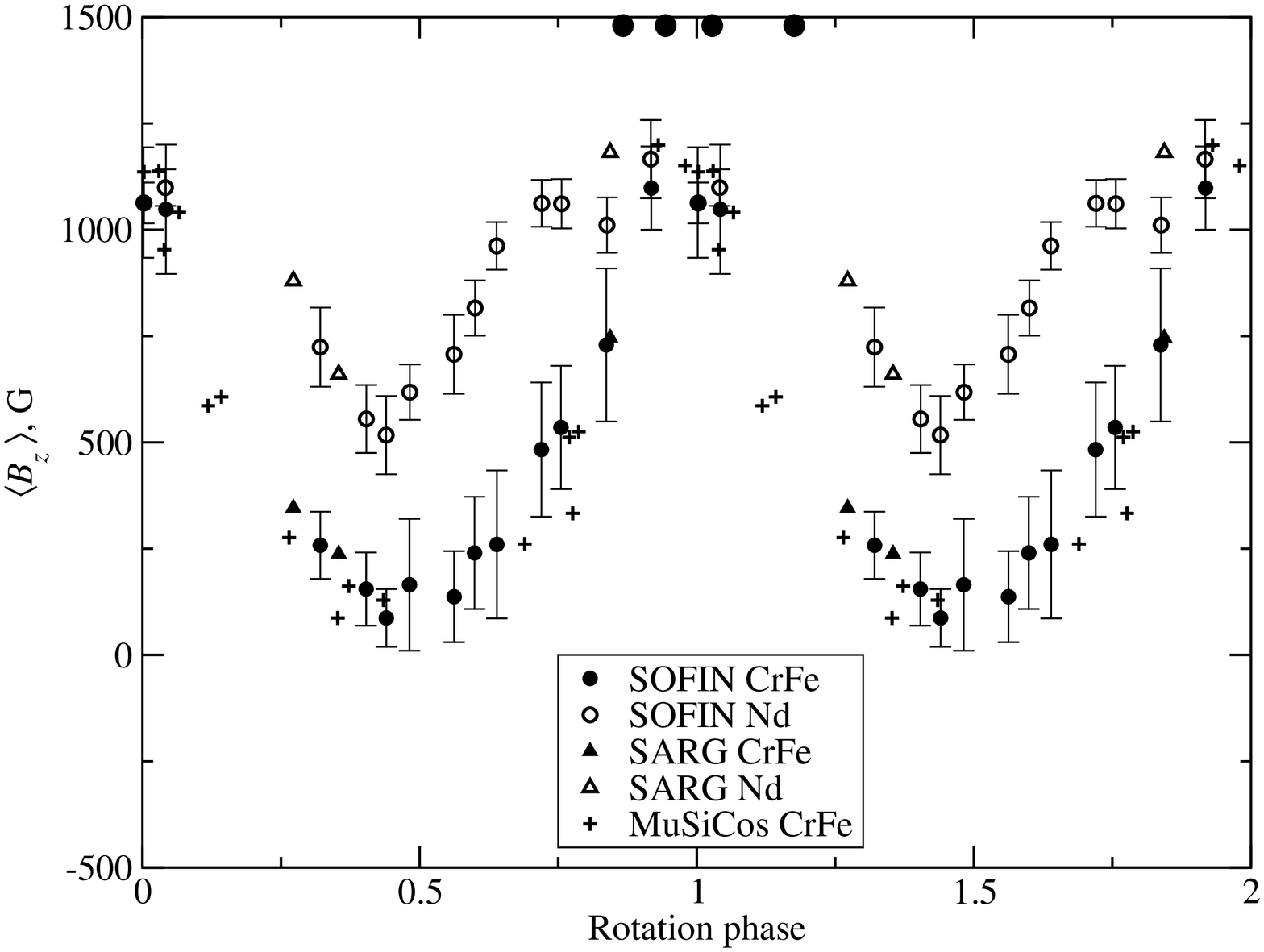}
\caption{\footnotesize Magnetic field variation during a rotation
cycle for different groups of elements. Large filled circles at
the top indicate the rotation phases of the time-series
spectroscopy.} \label{magnet}
\end{figure}

The longitudinal magnetic field \bz\ was measured with 7 Cr and Fe,
and 5 Nd\ii\ and Nd\iii\ lines. Fig.\,\ref{magnet} shows the
\bz\ variations with rotation phase and compares them with the
most recent measurements taken from the literature
(\citealt{LC04}, \citealt{RW05}). Obviously, for the two groups of
elements the magnetic curves are different. Magnetic Doppler
imaging of HD\,24712, which is in progress, should answer the
question whether this phenomenon is due to a different surface
distribution of the species in a star with dipolar magnetic field
geometry or due to a more complicated magnetic field geometry. Detailed
knowledge of the
latter is crucial for modelling pulsation in roAp stars.

\section{NLTE analysis of line formation depth and modelling pulsation
 in HD\,24712}

\begin{figure}[t]
\includegraphics[width=55mm, angle=-90]{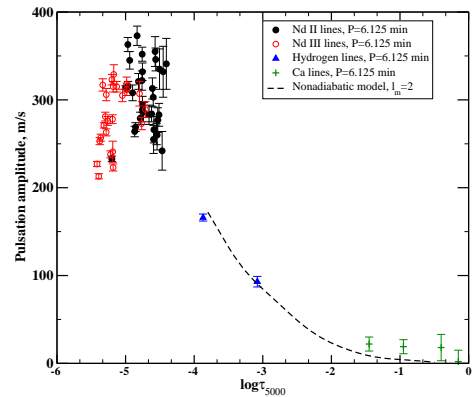}
\caption{ \footnotesize Pulsation amplitudes vs. depths of line
formation. } \label{depth}
\end{figure}

The depths of the line forming regions were calculated for Nd
based on a NLTE stratification analysis \citep{MRR05}, for
hydrogen on NLTE calculations (Mashonkina, private communication)
and for Ca based on a LTE stratification study.  The dependency of
the RV amplitudes from the optical depth is shown in
Fig.\,\ref{depth}. A preliminary model for nonadiabatic nonradial
pulsation was calculated for $M=1.65\,M_{\odot}$,
$\log(L/L_{\odot})$\,=\,0.912, \teff\,=\,7350\,K, and $R=1.77
R_{\odot}$ (see \citealt{Saio}, and this conference). The model
predicts the observed range of photometric frequencies as a
superposition of modes with $\ell$\,=\,1, 2, 3 for a polar
magnetic field of $B_p$\,$\approx$\,6\,kG, although exciting or
damping of these modes depends on the radiative transport
treatment. 
Calculated RV amplitudes as a function of depth are shown in Fig.\,\ref{depth}
(dashed line) for a quasi-quadrupole  mode with a frequency of 2.71\,mHz
(6.15 min).
The model is satisfactory and
predicts the existence of a `nodal layer' at $-0.15\,<\,\log
\tau_{5000}\,<\,-1.4 $ where RV amplitudes are close to zero and
where most of the spectral lines not showing measurable pulsation
amplitudes are formed. The model predicts also a luminosity
maximum at 0.38\,$P_{\rm puls}$ (2.4 radian) after the radius
minimum at $\log \tau_{5000}$\,=\,-3.7. The difference between the
observed and predicted phase lag is only 0.2 (1.2 radian) which
is small enough taking the preliminary character of the proposed
model into account. The decreasing observed RV amplitudes above
$\log \tau_{5000}$\,=\,-5 seem to be real and may be explained by
energy dissipation of the propagating wave.

\section{Conclusions}

Simultaneous photometric and spectroscopic observations of the
roAp star HD\,24712 allowed to design a consistent picture of
stellar pulsation from the envelope to the upper atmosphere which
is in good agreement with the proposed model of nonadiabatic
nonradial pulsation.

\begin{acknowledgements}

We are thankful to the MOST Science Team for providing us with the
photometric data and frequency analysis prior of publication. This
work was supported by the RFBR grants 04-02-16788 and 05-02-26664,
by the Austrian FFG-ALR (MOST Ground Station) and Science Fonds
(FWF-P17580N2).

\end{acknowledgements}

\bibliographystyle{aa}

\end{document}